\documentclass[final,journal]{IEEEtran}
\usepackage{amsmath}
\usepackage[utf8]{inputenc}
\usepackage{graphicx}
\usepackage{subfigure}
\usepackage{cite}

\begin{document}
\title{Simultaneous maximum-likelihood reconstruction for x-ray grating based phase-contrast tomography avoiding intermediate phase retrieval}
\author{André~Ritter$^1$, Florian~Bayer$^1$, J\"{u}rgen~Durst$^1$, Karl~G\"{o}del$^1$, Wilhelm~Haas$^{1,2}$, Thilo~Michel$^1$, Jens~Rieger$^1$, Thomas~Weber$^1$, Lukas~Wucherer$^1$ and Gisela~Anton$^1$%
\thanks{$^1$University of Erlangen, Erlangen Centre for Astroparticle Physics, Erwin-Rommel-Stra\ss e 1, 91058 Erlangen, Germany}%
\thanks{$^2$University of Erlangen, Pattern Recognition Lab, Martensstr. 3, 91058 Erlangen, Germany}}

\maketitle
\begin{abstract}
Phase-wrapping artifacts, statistical image noise and the need for a minimum amount of phase steps per projection limit the practicability of x-ray grating based phase-contrast tomography, when using filtered back projection reconstruction. For conventional x-ray computed tomography, the use of statistical iterative reconstruction algorithms has successfully reduced artifacts and statistical issues. In this work, an iterative reconstruction method for grating based phase-contrast tomography is presented. The method avoids the intermediate retrieval of absorption, differential phase and dark field projections. It directly reconstructs tomographic cross sections from phase stepping projections by the use of a forward projecting imaging model and an appropriate likelihood function. The likelihood function is then maximized with an iterative algorithm. The presented method is tested with tomographic data obtained through a wave field simulation of grating based phase-contrast tomography. The reconstruction results are in agreement with the expectations and proof the validity of the concept.
\end{abstract}
\IEEEpeerreviewmaketitle 

\section{Introduction}
\IEEEPARstart{X}{-ray} grating based phase-contrast imaging has been a focus of interest for several years \cite{wilkins96, david02, momose03, momose05, momose06, pfeiffer06, clauser1992, weitkamp05}. Tomography \cite{pfeiffer2007a} with this imaging method is possible but still suffers from unsolved problems. Phase-wrapping artifacts \cite{haas2011}, statistical requirements \cite{raupach2011, Weber2011} and the need for a minimum amount of three phase steps per projection impose limits on the practicability of the reconstruction using filtered back projection. For conventional x-ray computed tomography the use of statistical iterative reconstruction algorithms allows for a reduction of artifacts and statistical image noise. In this work, a reconstruction method for grating based phase-contrast tomography based on maximum likelihood estimation with an iterative algorithm is presented. The method avoids the intermediate retrieval of absorption, differential phase and dark field projections and directly reconstructs tomographic cross sections from projection data.

\section{Grating based phase-contrast imaging}
\label{sec:imaging}

\subsection{Interferometer setup}
\label{sec:interferometer}
\begin{figure}[htbp]
\begin{center}
\includegraphics[width=.49\textwidth]{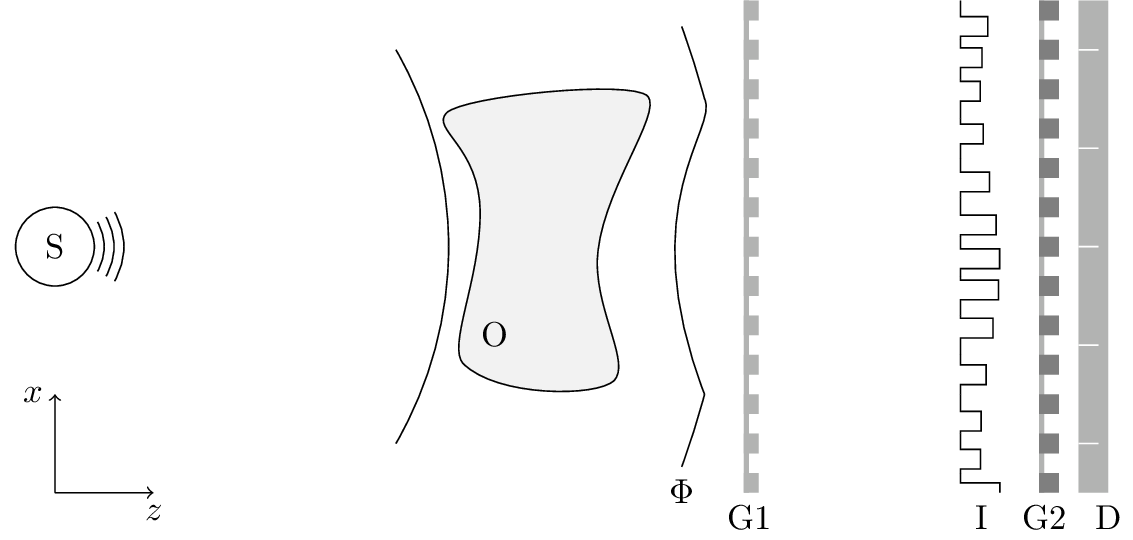}
\caption{Sketch of a Talbot interferometer setup for x-ray grating based phase-contrast tomography. It consists of an x-ray source S, a diffraction grating G1, an analyzer grating G2 and a pixelated x-ray detector D. An object O changes the amplitude and shifts the phase of the x-ray wave front $\Phi$. As a consequence the diffraction pattern I at grating G2 has slight modifications. The orientation of the grating bars is in y-direction.}
\label{fig:setup}
\end{center}
\end{figure}
For x-ray grating based phase-contrast tomography a Talbot interferometer \cite{momose03, clauser1992} is used as shown in figure \ref{fig:setup}. The interferometer consists of a diffraction grating G1 which is coherently illuminated by an x-ray source S. Downstream the diffraction grating a Talbot diffraction pattern can be observed. If the source does not provide illumination with sufficient spatial coherence, the altered Talbot-Lau \cite{Jahns1979} setup can be used where a source grating is positioned between x-ray source and diffraction grating, fulfilling certain conditions. Objects that are present in the x-ray wave field lead to local modifications of the diffraction pattern due to the absorption and the phase shift of the x-ray waves passing through the object.

\subsection{Phase stepping and retrieval}
\label{sec:phasestepping}
To detect the modifications of the diffraction pattern and obtain images, the phase stepping approach \cite{weitkamp05, morgan82, creath88} is used. An absorbing analyzer grating G2 is positioned parallel to G1 in a suitable distance. The period of the analyzer grating equals the period of the diffraction pattern generated by G1. The analyzer grating G2 is then shifted in several discrete steps in x-direction perpendicular to the grating bars. For each of these so called phase steps the intensity transmitted by the analyzer grating is recorded in each pixel of a pixelated x-ray detector D.  By using least square methods or discrete Fourier transform (DFT) algorithms the mean intensity, the phase and the visibility can be retrieved from the phase step intensity modulation in each pixel. The absorption, differential phase and dark field image can be obtained by comparing these values for a data set with an object in the beam path with the values from a reference data set without an object in the beam path.

\section{Method}
\label{sec:method}
For grating based phase-contrast tomography several phase step projections from different projection angles have to be acquired. In general absorption, differential phase and dark field projections are retrieved from the phase stepping projections as described in section \ref{sec:phasestepping}. Filtered back projection algorithms are then used to independently reconstruct tomographic cross sections for the absorption, differential phase and dark field projections. This has several drawbacks. To obtain absorption, differential phase and dark field images at least three phase step projections per projection angle have to be acquired. Phase wrapping that may occur in single pixels of a differential phase image can cause severe artifacts in reconstructed cross sections. Last but not least, the retrieved phase in a pixel tends to become arbitrary when photon statistics is not sufficient for that pixel. To avoid these drawbacks a tomographic reconstruction based on maximum likelihood methods \cite{barlow89} is proposed in this work.

\subsection{A likelihood function for differential phase-contrast tomography}
\label{sec:likelihood}
In the following the index $i$ denotes a pixel in a projection of the tomographic data while $s$ indicates the phase step. For every $i$ and every phase step $s$ we have a given measured intensity $N_{i,s}$. The aim of this approach is to reconstruct tomographic cross sections from the phase step projection data set $N$ without having to retrieve absorption, differential phase and dark field projections explicitly. Therefore, a model of the imaging process which provides the expected phase step intensitiy $\overline{N}_{i,s}$ has to be used. The expected phase step intensitiy $\overline{N}_{i,s}$ depends on the object parameters $\theta$ which include the distribution of the linear attenuation coefficient $\mu(\mathbf{r})$ and refractive index decrement $\delta(\mathbf{r})$ of the object. The likelihood of the object parameters $\theta$ for the given phase step projection data set $N$ is given by
\begin{equation}
\label{eqn:likelihood}
L\left(\theta\vert N\right)=\prod_{i,s}P_{\overline{N}_{i,s}(\theta)}\left(N_{i,s}\right),
\end{equation}
where $P_{\overline{N}_{i,s}(\theta)}\left(N_{i,s}\right)$ is the probability for detecting value $N_{i,s}$ when expecting $\overline{N}_{i,s}(\theta)$. If the detected values are Poisson distributed, which is true when using a photon counting detector, this probability is given by
\begin{equation}
\label{eqn:poisson}
P_{\overline{N}_{i,s}(\theta)}\left(N_{i,s}\right)=
\frac{{\overline{N}_{i,s}(\theta)}^{N_{i,s}}}{N_{i,s}!}
e^{-\overline{N}_{i,s}}.
\end{equation}
To obtain the best estimate for the object parameters the global maximum of the likelihood function $L$ or the log-likelihood function $\log L$ with respect to $\theta$ has to be found.

\subsection{Imaging model}
\label{sec:model}
In the following, an imaging model for a grating based phase-contrast setup is presented. The intensity $\overline{N}_{i,s}$ which is expected to be detected in a pixel $i$ for a certain phase step $s$ can be described as
\begin{equation}
\overline{N}_{i,s}=N^0_i\cdot T_i\cdot\left[1+V^0_i\cdot D_i\cdot\cos\left(\phi^0_{i,s}-\phi_i\right)\right],
\end{equation}
where $N^0_i$ is the expected mean intensity and $V^0_i$ the expected visibility of the phase stepping modulation when no object is present. $T_i$ is the transmission, $D_i$ the dark field and $\phi_i$ the differential phase. $\phi^0_{i,s}$ is the sum 
\begin{equation}
\phi^0_{i,s}=\phi^0_i+\phi_s
\end{equation}
of the starting phase $\phi^0_i$ of the modulation for a pixel $i$ while the phase steps $\phi_s$ are determined by the translation steps of G2. In general, $N^0_i$, $V^0_i$ and $\phi^0_{i}$ can be acquired by taking a reference image without object.

$T_i$, $D_i$ and $\phi_i$ are tied to object properties and can thus be used to gain image information about the object. The transmission  $T_i$ that is seen by a pixel $i$ is given by the line integral
\begin{equation}
\label{eqn:lineT}
T_i=\exp\left(-\int_{\text{ray}_i}\mu(\mathbf{r})ds\right),
\end{equation}
where $\text{ray}_i$ denotes the integration path of the line integral starting at the x-ray source focus and ending at the center of the pixel. 

The dark field $D_i$ describes the change in the visibility due to the presence of the object. Small angle scattering and substructures of the object that are not resolvable with the imaging setup are suggested to cause this change in visibility \cite{bech09,Yashiro_Terui_Kawabata_Momose_2010}. An object property $\sigma(\mathbf{r})$ is introduced that can be used to quantify the influence of the object on the dark field $D_i$ by a line integral similar to the transmission:
\begin{equation}
\label{eqn:lineD}
D_i=\exp\left(-\int_{\text{ray}_i}\sigma(\mathbf{r})ds\right).
\end{equation}
The decrease in visibility can be described by a line integral as long as there are no scattering processes involved that have a broad angular distribution of the scattered photons. For example, this would not be the case for Compton scattering. The differential phase $\phi_i$ is given by 
\begin{equation}
\label{eqn:linephi}
\phi_i=C_\text{geom}\cdot\frac{\partial}{\partial\xi}\int_{\text{ray}_i}\delta(\mathbf{r})ds,
\end{equation}
where $C_\text{geom}$ is a constant factor depending on the imaging geometry and $\frac{\partial}{\partial\xi}$ denotes the derivative perpendicular to the grating bars of the analyzer grating. In the case of figure \ref{fig:setup}, the $\xi$-direction is equal to the x-direction, but this changes for different projection angles when the setup is rotated around the tomographic axis.

In our model $\mu$, $\delta$ and $\sigma$ are the object parameters of this imaging method. For numerical means the continuous distribution of these object parameters has to be sampled on a discrete grid. In the following the index $j$ denotes a voxel of such a sampled distribution which describes the object. The equations (\ref{eqn:lineT}) and (\ref{eqn:lineD}) can then be expressed as linear equations:
\begin{equation}
\label{eqn:lineTD}
T_i=\exp\left(-\sum_{j}M^T_{i,j}\mu_j\right)
\end{equation}
and
\begin{equation}
\label{eqn:lineDD}
D_i=\exp\left(-\sum_{j}M^D_{i,j}\sigma_j\right).
\end{equation}
The coefficient matrices $M^T_{i,j}$ and $M^D_{i,j}$ are equal due to the same form of equations (\ref{eqn:lineT}) and (\ref{eqn:lineD}). To calculate $M^T_{i,j}$ and $M^D_{i,j}$ Siddon's algorithm\cite{Siddon_1985} or a similar algorithm can be used.

Equation (\ref{eqn:linephi}) can be approximated in a similar way by the difference of two line integrals 
\begin{equation}
\label{eqn:linephiapprox}
\phi_i\approx \frac{C_\text{geom}}{p}\cdot\left(\int_{\text{ray}^+_i}\delta(\mathbf{r})ds-\int_{\text{ray}^-_i}\delta(\mathbf{r})ds\right),
\end{equation}
where $\text{ray}^{\pm}_i$ denotes the integration path starting at the x-ray source and ending at the right or left border of the pixel and $p$ is the distance between the pixel borders. From equation (\ref{eqn:linephiapprox}) the linear equation 
\begin{equation}
\label{eqn:linephiD}
\phi_i=\sum_{j}M^\phi_{i,j}\delta_j,
\end{equation}
with the coefficient matrix $M^\phi_{i,j}$ can be calculated.

\section{Results}
\begin{figure}[htbp]
   \centering
   \subfigure[$\mu$]{\includegraphics[width=.45\textwidth]{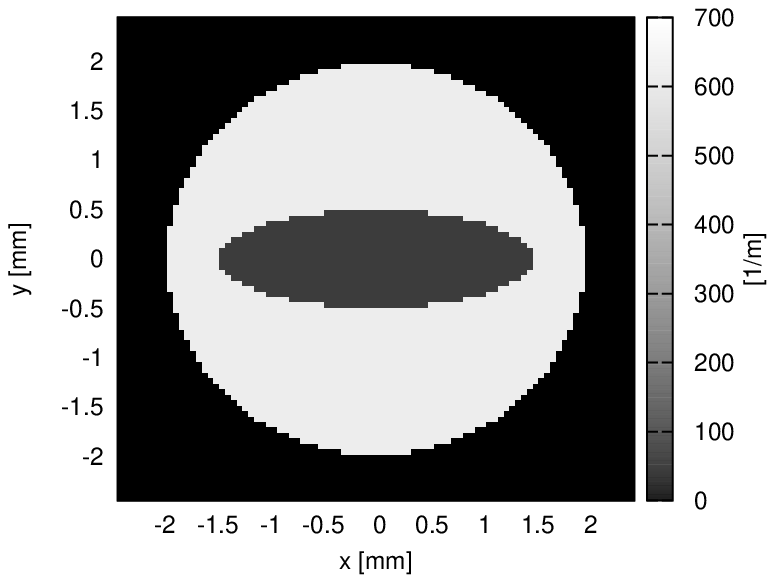}}
   \subfigure[$\delta$]{\includegraphics[width=.45\textwidth]{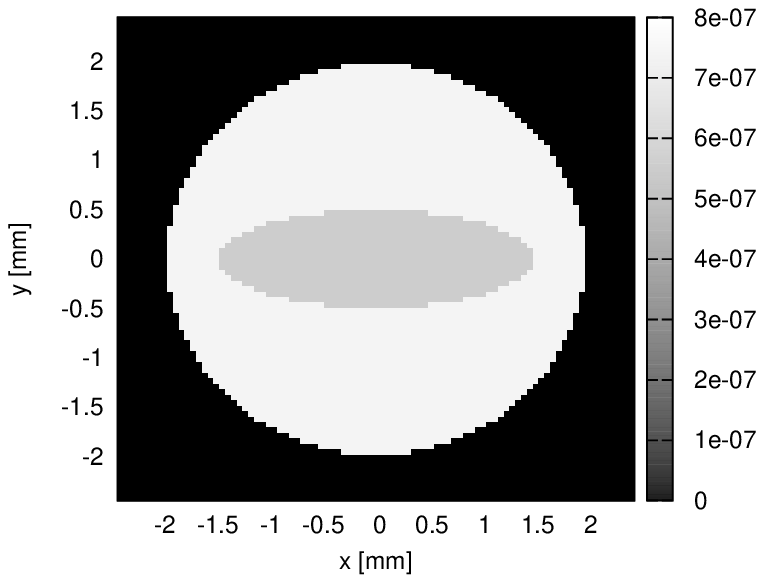}}   
   \caption{Distribution of (a) the attenuation coefficient $\mu$ and (b) the refractive index decrement $\delta$ of the circular object used for the simulation.}
   \label{fig:object}
\end{figure}
To proof the validity of the likelihood approach proposed in section \ref{sec:method} the tomography of a circular object with 2\,mm radius, shown in figure \ref{fig:object}, was simulated. The object consists of a high absorbing material with an elliptic insert of a low absorbing material. The tomography was done with 359 projections with four phase steps each. The parallel beam projections are distributed equally over 360 degrees. The simulated ideal photon counting detector has a pixel pitch of 55\,$\mu$m with 90 pixels and 100\% detection efficiency. The interferometer setup is designed for 20\,keV. The simulation is monochromatic with a photon energy of 20\,keV and $10^7$ incident photons per phase step and projection. The simulation was done with a wave field simulation code \cite{ritterdipl} using scalar diffraction theory.

\subsection{Iterative reconstruction}
To reconstruct the distributions of $\mu$, $\delta$ and $\sigma$ with the imaging model presented in section \ref{sec:model} a gradient ascent method was applied to the log likelihood function presented in section \ref{sec:likelihood}. Given some starting values $\mu^{(0)}$, $\delta^{(0)}$ and $\sigma^{(0)}$ the expected intensity $\overline{N}_{i,s}$ can be caclulated. Estimates of $\mu$, $\delta$ and $\sigma$ with a higher likelihood can be gained by iteratively using the update equations:
\begin{equation}
\label{eqn:update_mu}
\mu^{(n+1)}_j = \mu^{(n)}_j +\epsilon_\mu\cdot
\frac{\partial}{\partial \mu_j}\log L\left(\mu,\delta,\sigma\right)\vert_{\mu^{(n)},\delta^{(n)},\sigma^{(n)}},
\end{equation}
\begin{equation}
\label{eqn:update_delta}
\delta^{(n+1)}_j = \delta^{(n)}_j +\epsilon_\delta\cdot
\frac{\partial}{\partial \delta_j}\log L\left(\mu,\delta,\sigma\right)\vert_{\mu^{(n)},\delta^{(n)},\sigma^{(n)}},
\end{equation}
\begin{equation}
\label{eqn:update_sigma}
\sigma^{(n+1)}_j = \sigma^{(n)}_j +\epsilon_\sigma\cdot
\frac{\partial}{\partial \sigma_j}\log L\left(\mu,\delta,\sigma\right)\vert_{\mu^{(n)},\delta^{(n)},\sigma^{(n)}}.
\end{equation}
The parameters $\epsilon_\mu$, $\epsilon_\delta$ and $\epsilon_\sigma$ influence the convergence properties of the iterative reconstruction. If the values for these parameters are set too low, convergence will be slow resulting in a high amount of iterations needed to gain reconstructed images. If the values are too high convergence my not be possible at all. 

The derivatives in equations (\ref{eqn:update_mu}), (\ref{eqn:update_delta}) and (\ref{eqn:update_sigma})  are given by
\begin{equation}
\frac{\partial}{\partial \mu_j}\log L
=\sum_{i,s}\left(\overline{N}_{i,s}-N_{i,s}\right)\cdot M^T_{i,j},
\end{equation}
\begin{equation}
\frac{\partial}{\partial \delta_j}\log L
=\sum_{i,s}\left(1-\frac{N_{i,s}}{\overline{N}_{i,s}}\right)
\cdot N^0_iT_i\cdot V^0_iD_i\cdot \sin\left(\phi^0_{i,s}+\phi_i\right) \cdot M^\phi_{i,j}
\end{equation}
and
\begin{equation}
\frac{\partial}{\partial \sigma_j}\log L
=\sum_{i,s}\left(1-\frac{N_{i,s}}{\overline{N}_{i,s}}\right)
\cdot N^0_iT_i\cdot V^0_iD_i\cdot \cos\left(\phi^0_{i,s}+\phi_i\right) \cdot M^D_{i,j}
\end{equation}
with $N_{i,s}$ being the measured intensity. 

\subsection{Reconstruction results}
\begin{figure}[htbp]
   \centering
   \subfigure[$\mu$ reconstructed]{\includegraphics[width=.45\textwidth]{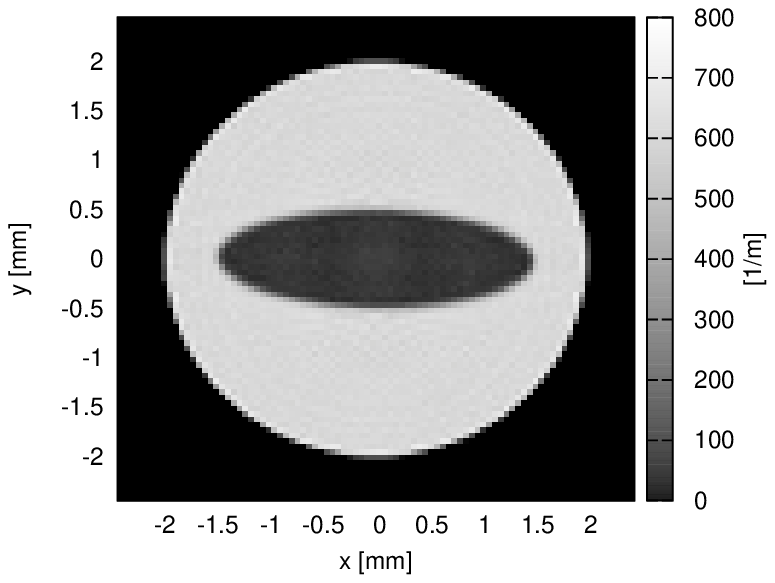}}
   \subfigure[$\mu$ difference]{\includegraphics[width=.45\textwidth]{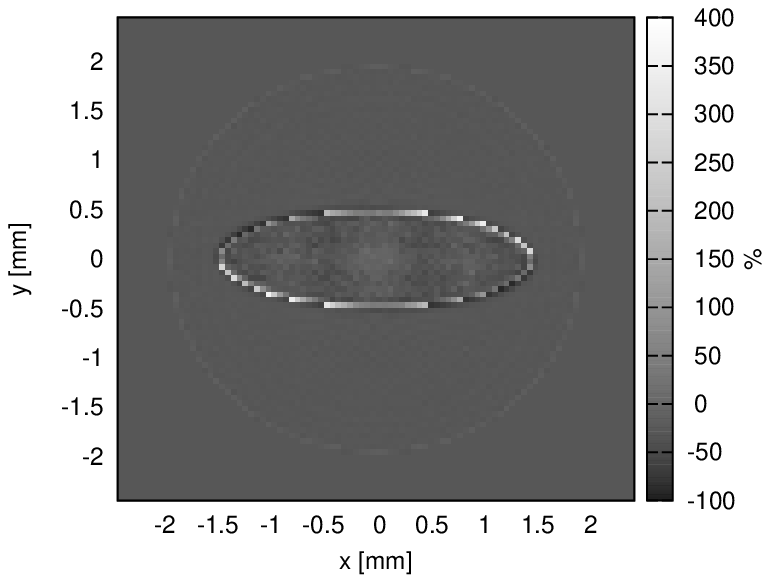}}   
   \caption{(a) Reconstructed image of $\mu$ after 100 iterations. (b) Relative difference of reconstructed $\mu$ compared to expected values.}
   \label{fig:iter100_mu}
\end{figure}
\begin{figure}[htbp]
   \centering
   \subfigure[$\delta$ reconstructed]{\includegraphics[width=.45\textwidth]{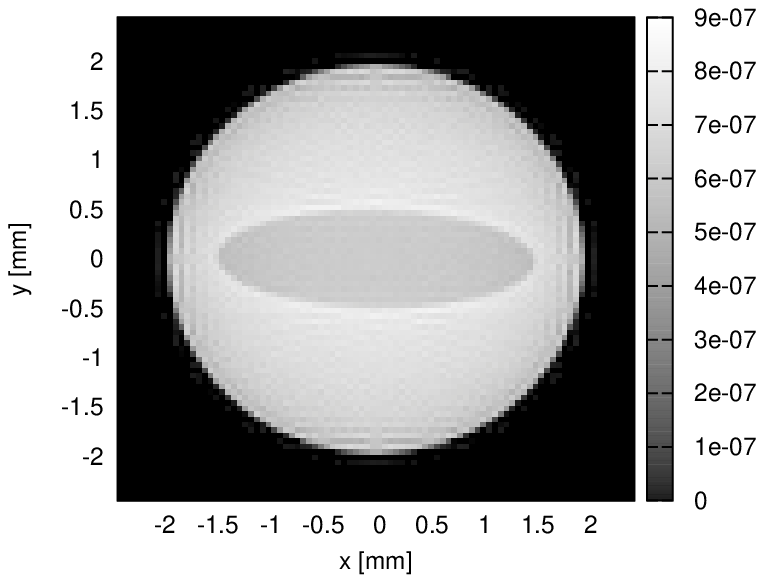}}
   \subfigure[$\delta$ difference]{\includegraphics[width=.45\textwidth]{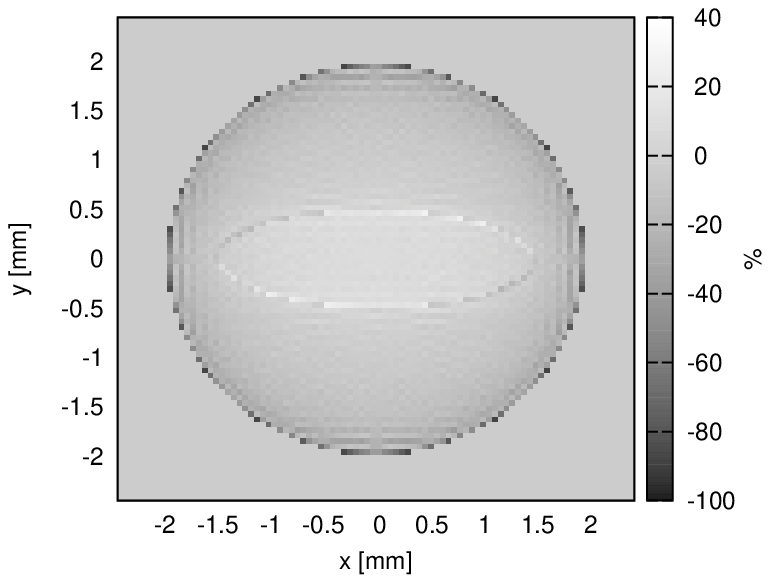}}   
   \caption{(a) Reconstructed image of $\delta$ after 100 iterations. (b) Relative difference of reconstructed $\delta$ compared to expected values.}
   \label{fig:iter100_delta}
\end{figure}
\begin{figure}[htbp]
   \centering
   \subfigure[$\sigma$ reconstructed]{\includegraphics[width=.45\textwidth]{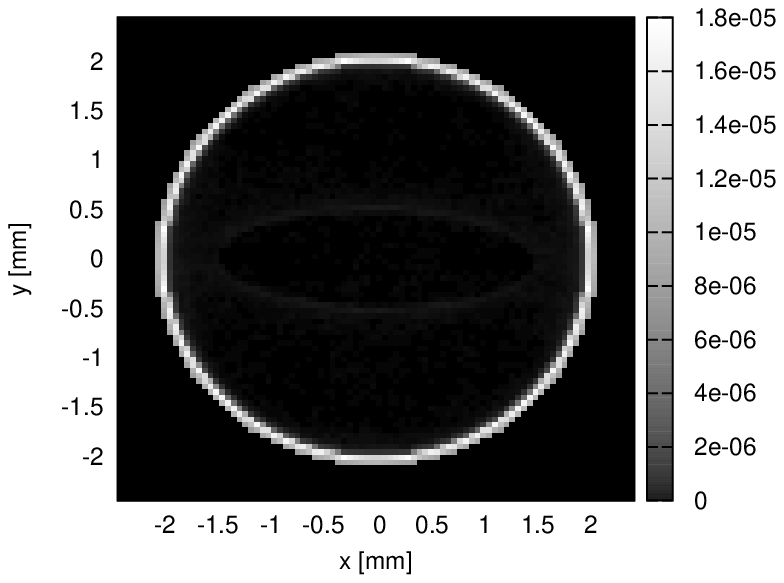}}
   \subfigure[$\sigma$ windowed]{\includegraphics[width=.45\textwidth]{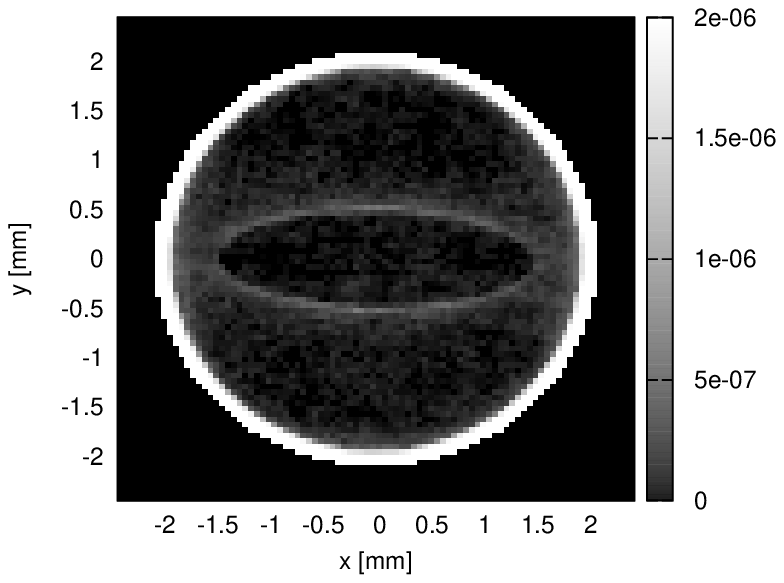}}
   \caption{Reconstructed images of $\sigma$ after 100 iterations. (a) Full range of $\sigma$. (b) Windowed range of $\sigma$.}
   \label{fig:iter100_sigma}
\end{figure}
The reconstruction was performed on a 90 by 90 matrix with a sampling distance of 55\,$\mu$m in both directions. The starting distributions of $\mu$ and $\delta$ was constructed in the following way. A circular area which is slightly bigger than the circular area of the expected reconstructed object was defined. This area was filled with a constant starting value within the order of magnitude which would be expected from the defined object. The remaining reconstructed region was set to zero. The starting distribution of $\sigma$ was set to zero for the whole reconstructed region. The iterative reconstruction was performed with $\epsilon_\mu=1\cdot10^0$, $\epsilon_\delta=2\cdot10^1$ and $\epsilon_\sigma=1\cdot10^{-8}$ and was aborted after 100 iterations.

Figure \ref{fig:iter100_mu} and figure \ref{fig:iter100_delta} show the reconstruction results for $\mu$ and $\delta$. In each figure the first subfigure (a) shows the result of the reconstruction. The second subfigure (b) shows the difference between reconstructed and expected values relative to the expected values. In the reconstruction of $\mu$, all features of the defined object are visible and in general the reconstructed values are in good agreement with the expected values. Differences are particularly visible at the edges and in the inner elliptic part of the object. The same features can also be seen in the reconstruction of $\delta$ with slightly higher relative differences between reconstruction and expectation. In figure \ref{fig:iter100_sigma}, subfigure (a) shows the reconstruction of $\sigma$, subfigure (b) shows the reconstructed values in a windowed range. In these figures only the edges of the object are visible. Altogether, it is apparent that the image reconstruction is possible with the described method.

\section{Conclusion}
In this work an iterative maximum likelihood reconstruction method for grating based phase-contrast tomography is presented. The reconstruction is able to directly reconstruct tomographic images from phase step projection data, without having to intermediately reconstruct absorption, phase and dark field projections on a per pixel basis. Furthermore, all cross-sectional images are obtained simultaneously. It is possible to reconstruct tomographic images from simulated phase stepping projections with this method. While there is still room for possible improvements in the method, the reconstructed images are in good agreement with the expectation.

The presented method might be able to overcome several problems which occur in grating based phase-contrast imaging. Phase wrapping artifacts can probably be  reduced or completely eliminated. Low statistics in single pixels might not be a problem as information can be gained from different pixels with better statistics. Last but not least, the mathematical principle of the reconstruction method has no requirement on the number of phase steps per projection, thus probably allowing reconstructions with less than three phase steps.

\section*{Acknowledgment}
This work was supported by the German Federal Ministry of Education and Research by the PHACT project (BMBF 01 EZ 0923 DLR).

\bibliographystyle{IEEEtran}
\bibliography{article}

\end{document}